 \documentclass[letterpaper, 10pt, conference]{ieeeconf}
 
 \IEEEoverridecommandlockouts                              

\makeatletter
\let\NAT@parse\undefined
\makeatother

\usepackage[T1]{fontenc}
\usepackage{url}
\usepackage[latin9]{inputenc}
\usepackage{array}
\usepackage{float}
\usepackage{units}
\usepackage{color}
\usepackage{graphicx}
\usepackage{multirow}
\usepackage{amsmath}
\usepackage{booktabs}

\usepackage{amsthm}
\usepackage{amssymb}
\usepackage{graphicx}
\usepackage{esint}
\usepackage[square, comma, sort&compress, numbers]{natbib}  
\usepackage{bbold}
\usepackage{stfloats} 
\makeatletter

\floatstyle{ruled}
\newfloat{algorithm}{tbp}{loa}
\providecommand{\algorithmname}{Algorithm}
\floatname{algorithm}{\protect\algorithmname}

\makeatother

\renewcommand{\t}{^{\mbox{\tiny\sf T}}}

\usepackage[english]{babel}

\makeatletter
\adddialect\l@ENGLISH\l@english
\makeatother

\makeatletter

\newcommand{\Rmnum}[1]{\expandafter\@slowromancap\romannumeral #1@}
\makeatother

\allowdisplaybreaks

\title {A Symbolic Regression Method for Dynamic Modeling \\ and Control 
of Quadrotor UAVs
}

\author{Wei Fang and Zhiyong Chen
\thanks{W. Fang is with the School of Information Science and Engineering, Central South University, Changsha, China.
Z. Chen is with the School of Electrical Engineering and Computing, The University of Newcastle,
Callaghan, NSW 2308, Australia. Z. Chen is the corresponding author 
{\tt\small  zhiyong.chen@newcastle.edu.au}. } 
}

\begin{document}

\maketitle

\thispagestyle{empty}

\begin{abstract}
This paper presents a mathematic dynamic model of a quadrotor unmanned aerial vehicle (QUAV) by using the symbolic regression approach  and then proposes a hierarchical control scheme for trajectory tracking. The symbolic regression approach is capable of constructing analytical quadrotor dynamic equations only through the collected data, which relieves the burden of first principle modeling. To facilitate position tracking control of a QUAV, the design of controller can be decomposed into two parts: a proportional-integral controller for the  position subsystem is first designed to obtain the desired horizontal position  and the backstepping method for the attitude subsystem is developed to ensure that the Euler angles and the altitude can fast converge to the reference values. The effectiveness is verified through experiments on a benchmark multicopter simulator. 

\end{abstract}




\section{Introduction}

The study of unmanned aerial vehicles (UAVs) has attracted considerable attentions due to their wide applications in many areas including formation control, disaster assistance, monitoring patrol, environmental detection, and so on. The main advantages of a quadrotor lie in its simple mechanical structure, capability of taking off and landing vertically and favorable maneuverability. However, its dynamic characteristics, due to strong coupling, underactuated property and nonlinearity, 
make designing a controller for position tracking become complicated.

 
An accurate model is a significant factor that determines the performance of a controller. 
First principle modeling and data driven modeling are the two main approaches for dynamic modeling;
see, e.g.,  \cite{2020Stability,2017Nonlinear,2013Modeling,2009Output,2016Learning,2006Real,2016Self}. The first principle modeling approach
uses physical principles to derive a dynamic model. In particular, it analyzes the forces applied
on the aircraft rotors and other modules, and derives the quadrotor dynamic equations utilizing mechanics and aerodynamics laws. For example, significant aerodynamic effects caused by the deformation of rotor blades were considered in \cite{2017Nonlinear}. The dynamic model constructed by this approach is capable of reflecting the flight behavior for more  situations. 
However, the derivation process and the analytical form of the dynamic equations 
are usually complicated and hence the controller design becomes challenging.

The other modeling approach is to identify a dynamic model 
based on the collected input excitation and output response data from a real system. 
It is worth nothing that the accuracy of an identified model particularly depends on the identification algorithms and the collected data.  Various identification techniques have been proposed in literature such as neural networks \cite{2009Output,2016Learning}, fuzzy-logic \cite{2006Real}, reinforcement learning \cite{2016Self}, and local linear regression \cite{2012Efficient}, etc. However, these methods usually suffer from drawbacks including local linearization (local linear regression), complicated calculation and low interpretability (neural networks). 
In particular, the purely data driven methods can accurately fit system dynamics, however, they only offer a black box model rather than 
analytical dynamic equations. Therefore, many advanced control theories based on analytical models cannot be used.

In this paper, we adopt symbolic regression to construct a dynamic quadrotor model, which can somehow overcome the aforementioned disadvantages. In particular, this data driven approach gives an analytical model 
that captures the physical characteristics of the quadrotor dynamics  and hence model based control theories can be applied.

Symbolic regression is an automated calculation approach that attempts to explore the inherent relationships only from  data and establishes an interpretable model. A model constructed by symbolic regression does not require a pre-specified  structure, which offers symbolic regression various potential applications \cite{Schmidt2009Distilling,2015Modeling,2019Multi,2019Influential,Vladislavleva2013Predicting,2019Modeling}. For instance,  a meta-heuristic algorithm was developed in
\cite{Schmidt2009Distilling} to distill experimental data captured from various physical systems into analytical natural laws, including Hamiltonians, Lagrangians and other laws of geometric. 
Symbolic regression was adopted to find potential relation under the oil production data and made predictions for the peak of oil production in \cite{2015Modeling}. The work in \cite{2019Multi}  discovered an underlying multi-parameter model of a system and that in \cite{2019Influential} revealed the important factors of the carbon emissions intensity using symbolic regression


In terms of tracking control for a quadrotor, researchers have presented various control techniques, such as dynamic inversion control \cite{Das2009Dynamic}, adaptive control \cite{4796260,2015Attitude}, backstepping control \cite{2016Robust}, sliding-mode control \cite{Luque2012Robust,2016Nonlinear}, LMI-based linear control \cite{2013LMI}, and so on. 
These controllers have their features. To the best of our knowledge,  few works have mentioned utilizing symbolic regression to construct an analytic model and a model-based control law. Therefore, we propose a tracking control scheme based on the analytic model obtained by symbolic regression.  

From the above discussion, the main contribution of this paper is to use symbolic regression to build an accurate analytic QUAV model, based on which a hierarchical  control scheme is designed for position tracking incorporating backstepping and dual-loop control techniques. Extensive
experimental results on a benchmark multicopter simulator demonstrate the effectiveness of the proposed symbolic regression based  model and the associated controller.

\section{Dynamic Modeling}\label{three_section}

In this section, we provide some background of a quadrotor dynamic model,  discuss how  symbolic regression
is used  for dynamic modeling, and then verify the effectiveness of the approach.  
The Matlab and PixHawk based multicopter simulator    developed by  
the BUAA Reliable Flight Control Group \cite{2017Introduction} is used throughout the paper as the benchmark experiment platform.

\subsection{A Quadrotor UAV Model}\label{subsection1}

\begin{figure}[htbp]
	\centering
	\includegraphics[scale=0.45]{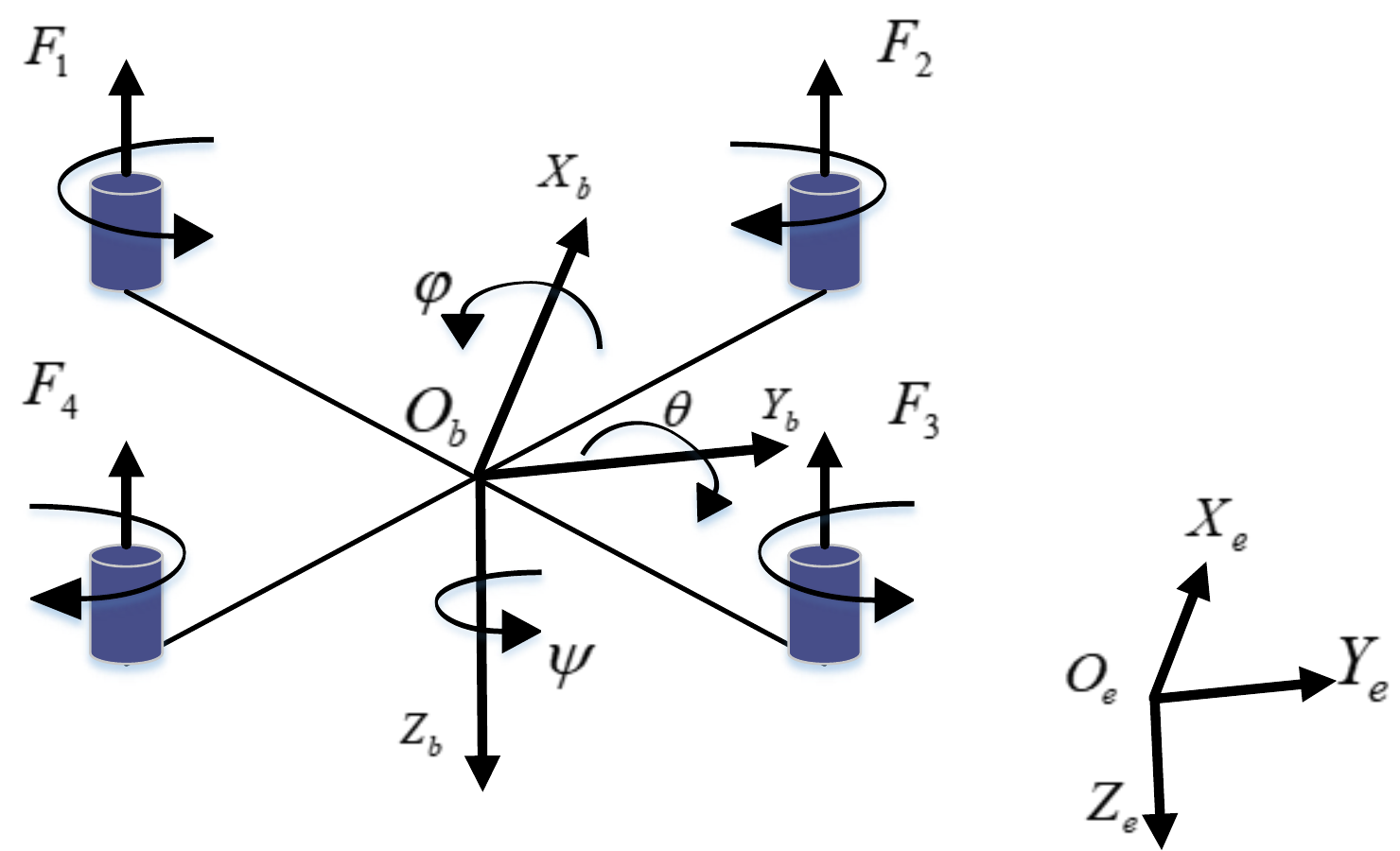}
	\caption{Schematic diagram of a quadrotor UAV}
	\label{fig_uav}
\end{figure}

A QUAV model represents a nonlinear system of four inputs and six degrees of freedom. It is composed of four individual electric motors and an X-shape airframe, as shown in Fig.~\ref{fig_uav}. The torque and thrust can be changed  by adjusting the rotation speed of each propeller, thereby further accomplishing different motions.

The dynamics of a quadrotor are explained using two reference frames. 
 The inertial frame $X_e$-$Y_e$-$Z_e$ is defined relative to  the ground, with the gravity pointing along the positive direction of the $Z_e$ inertial axis. Let the vector $[x,y,z]\t \in \mathbb{R}^{3}$ denote the position of the center of mass in the inertial frame. The velocity and Euler angles of the quadrotor with respect to the inertial frame are represented by the vector $[\dot{x},\dot{y},\dot{z}]\t \in \mathbb{R}^{3}$ and $[\phi,\theta,\psi]\t \in \mathbb{R}^{3}$, respectively.  The body frame $X_b$-$Y_b$-$Z_b$ is associated with the orientation of the quadrotor, with the thrust direction  always consistent with the negative direction of the $Z_b$ body axis. And the vector $[w_x,w_y,w_z]\t \in \mathbb{R}^{3}$ represents the angular velocity  in the body frame.
It is worthwhile pointing out that the roll and pitch angles are limited to ($-\pi/2,\pi/2$) and the yaw angle is limited to ($-\pi,\pi$),
which is physically meaningful. Then, the complete 12 dimensional state vector is denoted as \begin{align*}
\xi=  &\left[ \begin{array}{cccccccccccc}    
  x  &y & z& \dot x &\dot y &\dot z &\phi & \theta & \psi & w_x &w_y & w_z 
   \end{array} \right]\t.
\end{align*}

Let $\xi_i$, $i=1, \cdots, 12$, be the $i$-th element of the vector $\xi$. 
The control input $u_i$ represents the rotation speed of each motor, for $i=1,\cdots,4$.
Denote $u =\left[ \begin{array}{cccc}    
  u_1 & u_2& u_3& u_4
   \end{array} \right]\t$.
The dynamic model is of the following form
\begin{align}
\dot{\xi}_1 &=\xi_4 \nonumber \\
\dot{\xi}_2 &=\xi_5 \nonumber\\
\dot{\xi}_3 &=\xi_6 \nonumber\\
\dot{\xi}_i &=\zeta_i (\xi, u),\; i=4,\cdots, 12 \label{model}
\end{align}
with the some functions $\zeta_i (\xi, u),\; i=4,\cdots, 12$,  to be determined.


\subsection{Symbolic Regression Modeling}\label{subsection2}

Symbolic regression (SR)  is an established method that can effectively generate theories of causation in the form of mathematical equations from a collected data set. Unlike the traditional regression methods that require a fixed-form model derived from priori knowledge, SR automatically searches both the parameters and the form of equations. The expression of equations is composed of  common functions ($\sin$, $\cos$, $\sqrt{\cdot}$),  algebraic operators ($+$, $-$, $\times$, $\div$), constants and relative  variables. The selection of equations depends on its complexity and fitness. And the fitness function  $e(f^{*}(x),y)= \sum_{i=1}^n \lvert f^{*}(x^i)-y^i \rvert$ is widely used in SR, where $x$ denotes the input variable and $y$ the output variable, $n$ is the amount of pairs of input-output data $(x^i, y^i)$, and $f^{*}$ is the equation learned by SR. Since the  computation complexity caused by a large search space, genetic programming (GP)  \cite{2014Balancing} is typically implemented in SR.     
 GP is a meta-heuristic algorithm that adopts the principle of biological evolution to update candidate solutions.

 In the initial phase of modeling, we generate multiple groups of random  propeller rotation speeds within different ranges as
 the input data set. Then, by applying the input data set on the UAV simulator, we collect the output data set 
 which includes velocity, acceleration, angular velocity, and angular acceleration.
To perform symbolic regression modeling, we import the input and output data into the Eureqa software, select the formula building-blocks, e.g., $\left \{{\rm constant},+,-,\times, \div,\sin,\cos \right\}$ and the absolute error metric for training. As a result, the analytic equations through efficient training and sifting  are obtained as follows,
\begin{align*}
\zeta_4 (\xi, u) =& -0.00768 u_1 \cdot \sin(\xi_7) \cdot \sin(\xi_9)-1.63e^{-5} u_2 \cdot u_4 \\ & 
 \cdot \sin(\xi_7) \cdot \sin(\xi_9)-0.00665 \xi_8 \cdot u_1 \cdot \cos(\xi_7) \\
& \cdot \cos(\xi_9)  -1.63e^{-5} \xi_8 \cdot u_2 \cdot u_4 \cdot \cos(\xi_7) \cdot \cos(\xi_9)\\	
\zeta_5 (\xi, u) = &7.87e^{-6} u_1^2 \cdot \sin(\xi_7+\xi_9)+7.78e^{-6} u_2^2 \\ & \cdot \sin(\xi_7+\xi_9) +7.78e^{-6} u_3^2 \cdot \sin(\xi_7+\xi_9) \\ & +7.78e^{-6} u_4^2  \cdot \sin(\xi_7+\xi_9) -0.0441\\
\zeta_6 (\xi, u) = & 9.44 -6.90 e^{-6}  \cos(\xi_7)\cdot (u_1^2+u_2^2+u_3^2+u_4^2) \\ 
& -0.000289  \xi_6^2 \\
\zeta_7 (\xi, u)  = &\sin(\xi_7)\cdot \frac{\sin(\xi_8)}{\cos(\xi_8)} \cdot \xi_{11}+\cos(\xi_7) \cdot \frac{\sin(\xi_8)}{\cos(\xi_8)}  \cdot \xi_{12} \\& + \xi_{10} \\
\zeta_8 (\xi, u)  = & \cos(\xi_7)\cdot \xi_{11}-\sin(\xi_7) \cdot \xi_{12} \\
\zeta_9 (\xi, u)  = & \frac{\sin(\xi_7) }{ \cos(\xi_8)} \cdot \xi_{11}+\frac{\cos(\xi_7) }{ \cos(\xi_8)} \cdot \xi_{12}\\
\zeta_{10} (\xi, u)  = & 0.697 \xi_{11} \cdot \xi_{12}+8.33  e^{-5}  (u_2^2+u_3^2-u_1^2-u_4^2)\\
\zeta_{11} (\xi, u)  = & -0.708  \xi_{10} \cdot \xi_{12}+8.03  e^{-5}  (u_1^2+u_3^2-u_2^2-u_4^2)\\
\zeta_{12} (\xi, u)  = & 0.0219  \xi_{10} \cdot \xi_{11}+4.86  e^{-6}  (u_1^2+u_2^2-u_3^2-u_4^2).
\end{align*}

In particular, the functions $\zeta_i (\xi, u)$, $i=7,8,9$, represent the conversion relationship between $\{\dot \xi_7,\dot \xi_8,\dot \xi_9\}$ and $\{\xi_{10},\xi_{11},\xi_{12}\}$, which are considered as known transform equations. The remaining functions  are obtained from different training data via symbolic regression.
%
%
For instance, in order to acquire the function $\zeta_4 (\xi, u)$, we generate four groups of training data (2,000 samples) using the inputs in different ranges of  0-300rad/s, 0-700rad/s, 0-800rad/s and 0-1000rad/s. Both model validation and controller design are  implemented on the equations learned from different training data, and the equation that performs the best  is selected in the final model.  The other functions in \eqref{model} are obtained in a similar way. 
It is worth mentioning that the propeller speeds for $u_i,i=1,\cdots,4,$ are constrained between 0 and 1000rad/s.

\subsection{Model Validation}\label{subsection3}   

\begin{figure}[htbp]
	\centering
	\includegraphics[width=8.5cm]{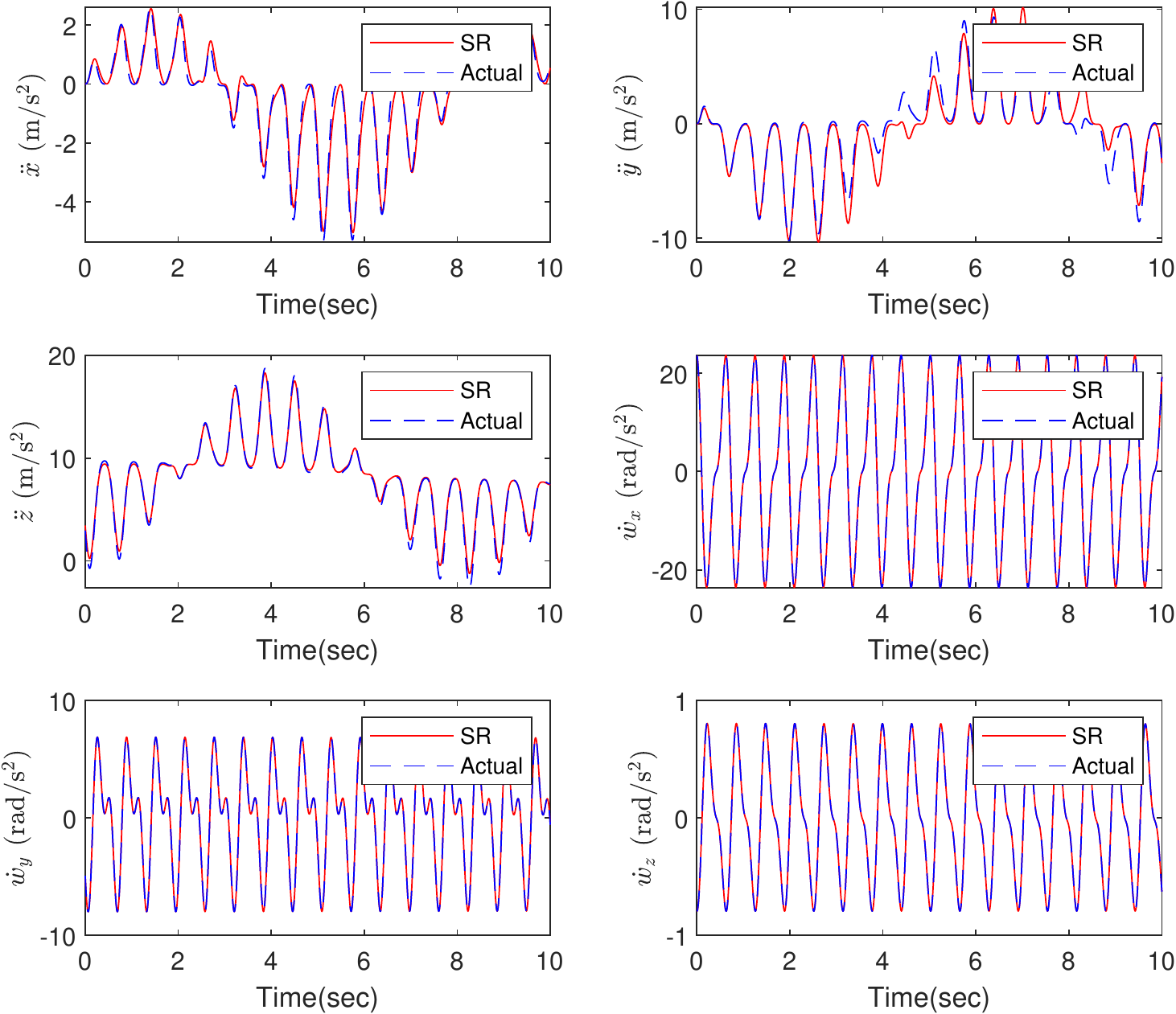}
	\caption{Comparison of the SR model with the actual system}
	\label{Variables_comparison}
\end{figure}

To validate the learned dynamic model, a test data set is generated using the following excitation inputs for 10 seconds,
	\begin{align*} 
 u_{1}=&300+300\sin(10t ) \\
 u_{2}=&300+300\sin(10t+\pi/4)\\
 u_{3}=&300+300\sin(10t+\pi/3)\\
 u_{4}=&300+300\sin(10t+\pi/6).
\end{align*}
 The reliability of the model is evaluated by comparing the six target variables ($\ddot{x},\ddot{y},\ddot{z},\dot{w}_x,\dot{w}_y,\dot{w}_z)$ obtained from the actual system and the SR model, respectively, and the result is shown in Fig.~\ref{Variables_comparison}.

Generally speaking, symbolic regression can discover the accurate model equations beneath data, especially for angular acceleration $\dot{w_x},\dot{w_y},\dot{w_z}$, which indicates that  symbolic regression is an effective modeling approach. Moreover, the RMSE, MAE and $R^2$ indexes are applied to quantitively reflect the approximation ability of SR;  the results are summarized in Table~\ref{table_test}. The accuracy of the analytic model is negatively correlated with the values of RMSE, MAE and $|1-R^2|$. 
 
\begin{table}[htbp]  
	 \caption{\label{tab:test}Testing errors for the learned SR model}   
	 \centering
	 \begin{tabular}{llll}
	 \toprule
	 Variable & RMSE & MAE &  $R^2$  \\
	 \midrule
	  \quad $\ddot x$ & 0.2632 & 0.2053 & 0.9707 \\
	  \quad $\ddot y$ & 1.2655 & 0.8021 & 0.8781 \\
	  \quad $\ddot z$ & 0.4011 & 0.2745 & 0.9907 \\
	  \quad $\dot w_x$ & 0.0034 & 0.0028 &1.0000\\
	  \quad $\dot w_y$ & 0.0013 & 0.0011 &1.0000\\
	  \quad $\dot w_z$ &$6.7079e^{-5}$ & $5.7108 e^{-5}$ &1.0000\\
	 \bottomrule
	 \end{tabular} 
	 \label{table_test}
  \end{table}

\section{Control Design Scheme}\label{four_section}
In this section, we focus on the controller design for position and attitude tracking simutaneously. 
However,  due to the underactuated property of a quadrotor, it is difficult to directly control  the six state variables $\left \{x,y,z,\phi,\theta,\psi \right \}$ simultaneously. Therefore, the state variables to be controlled are chosen as $x,y,z$ and $\psi$.

Let $x_d,y_d, z_d, \psi_d$ be the desired trajectories for $x, y, z, \psi$, respectively.   
With
\begin{align*}
e_x =x_d-x ,\; e_y = y_d-y,\; e_z = z_d-z, e_{\psi} =\psi_d-\psi
 \end{align*}
 the control target is to achieve the following asymptotic tracking behaviors
 \begin{align}
\lim_{t\rightarrow \infty}e_x(t) =0 ,\; 
\lim_{t\rightarrow \infty}e_y(t) =0 \label{limexy}
 \end{align}
 and 
  \begin{align}
\lim_{t\rightarrow \infty}e_z(t) =0 ,\; 
\lim_{t\rightarrow \infty}e_{\psi}(t) =0.\label{limezpsi}
 \end{align}

\begin{figure}[htbp]
	\centering
	\includegraphics[scale=0.3]{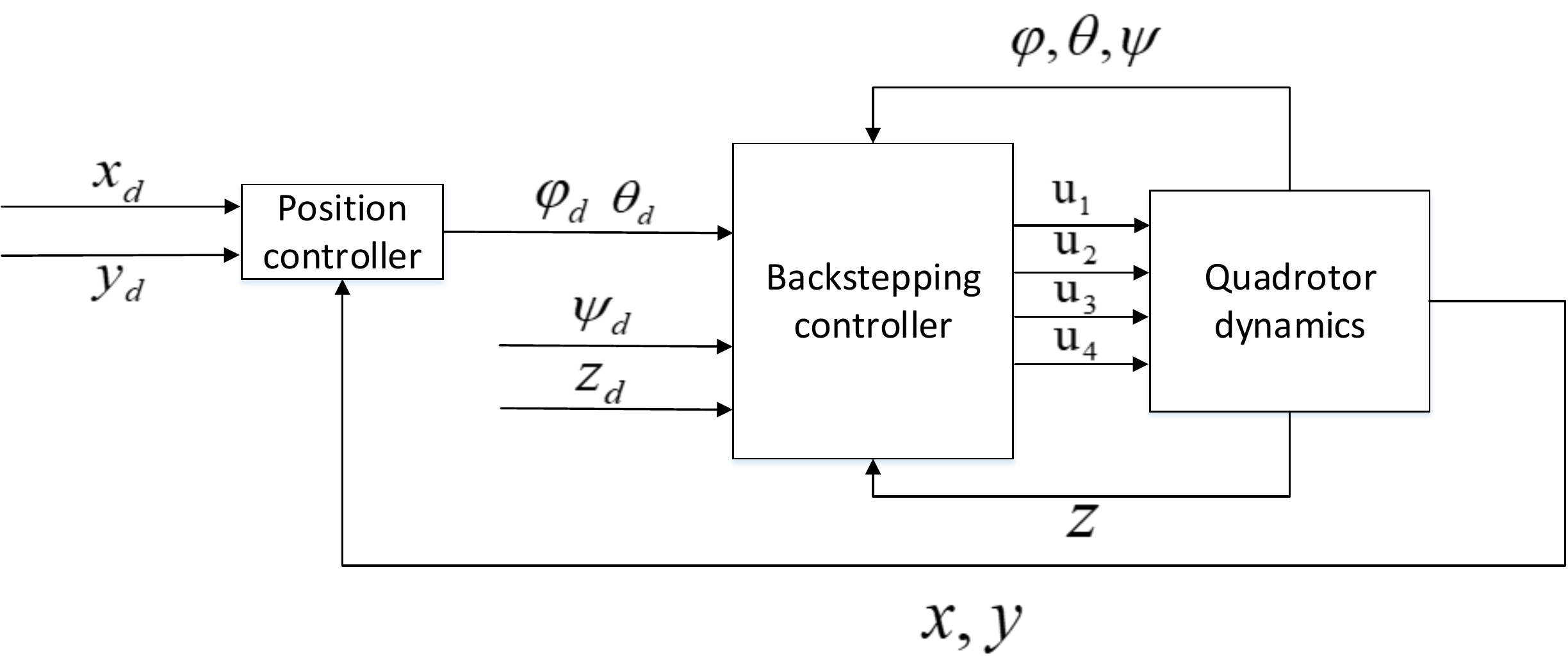} 
	\caption{Structure of the overall control system}
	\label{fig_framework}
\end{figure}

Given the reference position trajectory $x_d,y_d,z_d$ and the attitude angle $\psi_d$, we can derive the reference attitude angles $\phi_d,\theta_d$ and hence the control laws for $u_1,u_2,u_3,u_4$ from a backstepping control scheme.  
To avoid complex calculation, we employ a hierarchical control strategy consisting of the outer-loop control for horizontal position as well as the inner-loop control for attitude and altitude. The block diagram of the overall control system is shown in Fig~\ref{fig_framework}.
\subsection{Outer-loop controller design}\label{subsection1}

The objective of this part is to design a proportional-integral  (PI) controller for the outer loop subsystem to ensure the roll  and pitch angles  to be automatically tuned to drive the quadrotor to the desired horizontal position $x_d,y_d$. Based on \eqref{model}, the horizontal position subsystem  is described as follows:
\begin{align}
\dot{\xi}_1 &=\xi_4 \nonumber \\
\dot{\xi}_2 &=\xi_5 \nonumber\\
 \dot{\xi}_4&=\zeta_4 (\xi_7, \xi_8, \xi_9, u) \nonumber\\
\dot{\xi}_5&=\zeta_5(\xi_7,  \xi_9, u) \label{sys1}
\end{align}
where $\xi_7=\phi$ and $\xi_8=\theta$ are regarded as the virtual controls to this subsystem.

More specifically, we aim to solve the desired angle trajectories for $\xi_7$ and $\xi_8$, 
i.e., $\xi_{7d}=\phi_d$ and $\xi_{8d}= \theta_d$, from the following equations 
\begin{align}
 \zeta_4 (\xi_{7d}, \xi_{8d}, \xi_9, u) =
  k_{px} \dot e_x+k_{ix}  e_x+  \ddot x_d \nonumber\\
 \zeta_5(\xi_{7d},  \xi_9, u)=
  k_{py} \dot e_y+k_{iy}  e_y+  \ddot y_d.  \label{xi78d}
\end{align}
 The solutions to \eqref{xi78d} can be explicitly expressed as follows
\begin{align}
\xi_{7d}=&   \sin^{-1}((k_{py} \dot e_y+k_{iy}  e_y+  \ddot y_d  +0.0441)\nonumber \\
& /(7.87e^{-6}u_1^2+7.78e^{-6}u_2^2+7.78e^{-6}u_3^2 \nonumber\\ & +7.78e^{-6}u_4^2))-\xi_9 \nonumber\\
\xi_{8d}=& (k_{px} \dot e_x+k_{ix}  e_x+  \ddot x_d +0.00768 u_1 \nonumber\\
& \cdot \sin(\xi_{7d}) \cdot \sin(\xi_9)+1.63e^{-5}u_2 
 \cdot u_4  \nonumber\\ & \cdot  \sin(\xi_{7d}) \cdot \sin(\xi_9))/(-0.00665 u_1 \cdot \cos(\xi_{7d}) \nonumber\\
&    \cdot \cos(\xi_9)-1.63e^{-5}  u_2 \cdot u_4 \cdot \cos(\xi_{7d}) \cdot \cos(\xi_9)). \label{xi78d}
\end{align}
 
When $\xi_7$ and $\xi_8$ achieve $\xi_{7d}$ and $\xi_{8d}$, respectively, (to be elaborated
in the next subsection), the closed-loop system  \eqref{sys1} with 
$\xi_7 =\xi_{7d} $ and $\xi_8=\xi_{8d}$ becomes
\begin{align}
0 &= \ddot e_x + k_{px} \dot e_x+k_{ix}  e_x 
  \nonumber\\
  0 &= \ddot e_y + k_{py} \dot e_y+k_{iy}  e_y 
\end{align}
With proper selection of the parameters $k_{px}$, $k_{ix}$
$k_{py}$, and $k_{iy}$, the control objective \eqref{limexy}
can be achieved. 

The design of $\xi_{7d}$ and $\xi_{8d}$, as a virtual controller to 
the system \eqref{sys1}, is of a PI structure containing the 
proportional terms $k_{px} \dot e_x$ and  $k_{py} \dot e_y$
and the integral terms $k_{ix}  e_x  = k_{ix}\int \dot e_x$ and 
$k_{iy}  e_y  = k_{iy}\int \dot e_y$, as well as  feedforward compensation.

\subsection{Inner-loop controller design}\label{subsection2}

In this part, a backstepping controller  is designed to ensure 
$\xi_7=\phi$ and $\xi_8=\theta$ achieve $\xi_{7d}=\phi_d$ and $\xi_{8d}=\theta_d$, respectively, for the objective  \eqref{limexy}
as explained in the previous subsection, that is, 
 \begin{align}
\lim_{t\rightarrow \infty}e_\phi(t) =0 ,\; 
\lim_{t\rightarrow \infty}e_{\theta}(t) =0\label{limephitheta}
 \end{align}
for  \begin{align*}
e_\phi =\phi_d-\phi ,\; e_\theta = \theta_d-\theta.
 \end{align*} 
Also, it is to ensure the objective  \eqref{limezpsi}.
In other words, this subsection is concerned about the subsystem governing  
\begin{align}
 &\left[ \begin{array}{cccccccc}    
\xi_3 & \xi_6 & \xi_7  &  \xi_8 &  \xi_9  &  \xi_{10}  &   \xi_{11}  &  \xi_{12}
\end{array} \right]\t \nonumber \\
= &\left[ \begin{array}{cccccccc}    
z & \dot z &  \phi & \theta & \psi &w_x &w_y & w_z  
\end{array} \right]\t \label{state2}
\end{align} 
and the control objectives are \eqref{limezpsi} and \eqref{limephitheta}
where $z_d, \psi_d$ are the desired trajectories and
$\phi_d$ and $\theta_d$ designed in  \eqref{xi78d}.

 We introduce the following coordinate transformation 
  \begin{equation}
\begin{bmatrix}
\dot \phi \\
\dot \theta\\
\dot \psi \\
\end{bmatrix}
=
R
\begin{bmatrix}
	w_x \\
	w_y \\
	w_z \\
\end{bmatrix}
\end{equation}
with
\begin{equation*}
R=
\begin{bmatrix}
	1  & \frac{\sin(\phi)\sin(\theta)}{\cos(\theta)} & \frac{\cos(\phi)\sin(\theta)}{\cos(\theta)} \\
	0  & \cos(\phi) & -\sin(\phi) \\
	0  & \frac{\sin(\phi)}{\cos(\theta)} & \frac{\cos(\phi)}{\cos(\theta)}
\end{bmatrix},
\end{equation*}
then the variables $[\ddot \phi,\ddot \theta, \ddot \psi]^T$ can be obtained as follows:
\begin{equation}
	\begin{bmatrix}
		\ddot \phi \\
		\ddot \theta\\
		\ddot \psi \\
	\end{bmatrix}
	=
	R
	\begin{bmatrix}
	\dot w_x \\
	\dot w_y \\
	\dot w_z \\
	\end{bmatrix}
+ \dot R
	\begin{bmatrix}
	 w_x \\
	 w_y \\
	 w_z \\
\end{bmatrix}.
\end{equation}

Now, the subsystem governing the states \eqref{state2} can be equivalently transformed 
to a systems governing the following new state  vector 
\begin{align}
s 
=&\left[ \begin{array}{cccccccc}    
z & \phi & \theta & \psi & \dot z & \dot\phi & \dot \theta & \dot \psi  
\end{array} \right]\t.
\end{align} 
Let $s_i$ be the $i$-th element of $s$, for $i=1,\cdots, 8$.
Now, the dynamics for $s$ can be reorganized as follows
	\begin{align}
   \dot s_1 &=s_5 \nonumber\\
   \dot s_2 &=s_6  \nonumber\\
   \dot s_3 &=s_7 \nonumber \\
   \dot s_4 &=s_8 \nonumber \\
   \dot s_i &= \gamma_i (s, f) ,\; i=5, \cdots, 8 \label{ssystem}
	\end{align}
 where
	\begin{align*}
     \gamma_5 (s, f)= & 9.44-6.9e^{-6}  \cos(s_2) \cdot f_1 -0.000289 s_5^2 
          	\end{align*}
and $\gamma_i (s, f),\; i=6,7,8$ can be derived in a similar manner (with the details omitted).  
%
 The vector $f =\left[ \begin{array}{cccc}    
  f_1 & f_2& f_3& f_4
   \end{array} \right]\t$ represents the new control variables defined as follows
 		\begin{align}
		&f_1=u_1^2+u_2^2+u_3^2+u_4^2 \nonumber\\
		&f_2=u_2^2+u_3^2-u_1^2-u_4^2 \nonumber\\
		&f_3=u_1^2+u_3^2-u_2^2-u_4^2 \nonumber\\
		&f_4=u_1^2+u_2^2-u_3^2-u_4^2. \label{fu}
	\end{align}
 Let \begin{align*}
& s_{1d} =z_d,\;s_{2d} =\phi_d,\;s_{3d} =\theta_d,\;s_{4d} =\psi_d \\
 & e_i = s_{id} -s_i,\; i=1,\cdots,4. 
 \end{align*}
Now, the objectives  \eqref{limezpsi} and \eqref{limephitheta} are converted to 
 design of $f_i\; i=1,\cdots,4$ for \eqref{ssystem} such that 
 \begin{align}
\lim_{t\rightarrow \infty}e_i(t) =0, \; i=1,\cdots,4. \label{limei}
 \end{align}

The design follows a backstepping procedure, see, e.g., \cite{chen2002global}. First, we consider 
the Lyapunov function candidates
\begin{equation}
	V_i(e_i)=\frac{1}{2}e_i^2,
	\; i=1,\cdots, 4.
\end{equation}
The derivative of $V_i(e_i)$ along the trajectory of \eqref{ssystem}
satisfies  
\begin{equation}
	\dot V_i(e_i) =e_i \dot e_i=e_i (\dot s_{id}-s_{i+4}).
\end{equation}
Here, $s_{i+4}$ is the virtual control whose desired value is
\begin{equation}
	s_{(i+4) d}=\dot s_{i d}+c_i e_i.
\end{equation}
Let   
\begin{equation}
	\delta_i=s_{(i+4) d}-s_{i+4}.
\end{equation}
Now, we consider the  Lyapunov function candidates involving  both $e_i$ and  $\delta_d$ as follows
\begin{equation}
	W_i (e_i,\delta_i)=\frac{1}{2}e_i^2+\frac{1}{2}\delta _i^2,\; i=1,\cdots, 4,
\end{equation}
whose time derivative satisfies
	\begin{align*}
	&\dot W_i (e_i,\delta_i)\\ = & e_i \dot e_i+\delta_i \dot \delta_i\\
	= & e_i (\dot s_{id}-s_{i+4}) +\delta_i (\dot s_{(i+4) d}-\dot s_{i+4} )\\
	= & e_i (s_{(i+4) d}-c_i e_i -s_{i+4}) +\delta_i (\dot s_{(i+4) d}-\gamma_{i+4}(s,f) )\\
	= & e_i (\delta_i -c_i e_i ) +\delta_i (\dot s_{(i+4) d}-\gamma_{i+4}(s,f) )\\
	= & -c_i e_i^2 +\delta_i (e_i + \dot s_{(i+4) d}-\gamma_{i+4}(s,f) ). 
	\end{align*}
Letting
\begin{align}
 \gamma_{i+4} (s, f) = &e_i+\dot s_{(i+4)d}+c_{i+4}\delta _i \nonumber\\
=& s_{id}-s_i +\ddot s_{id}+c_i (\dot s_{id}-s_{i+4}) \nonumber\\
& +c_{i+4} (\dot s_{id}+c_i(s_{id}-s_i)-s_{i+4}) \label{fdesign}
\end{align}
gives
\begin{align*}
 \dot W_i (e_i,\delta_i) 
	=  -c_i e_i^2 -c_{i+4}\delta_i^2. 
 \end{align*}
If the two parameters $c_i$ and $c_{i+4}$ are selected as positive constants, one has the target
\eqref{limei} achieved.

 The control input $f$ can be solved from the equation \eqref{fdesign} for $i=1,\cdots, 4$. 
 For example, the input $f_1$  which stabilizes the altitude subsystem is thus presented as follows
 \begin{align*}
	f_1  = & (s_{1d}-s_1+\ddot s_{1d}+c_1(\dot s_{1d}-s_5)+c_5(\dot s_{1d} \\ 
& +c_1(s_{1d}-s_1)-s_5)-9.44+0.000289s_5^2) \\
& /(-6.9e^{-6}\cos(s_2)).
\end{align*}
 The inputs $f_2, f_3, f_4$ can be solved in a similar manner. 
And the control laws $u_1,u_2,u_3,u_4$ can be solved 
from \eqref{fu}.

%

\section{Numerical Experiments}\label{five_section}

In this section, the feasibility of symbolic regression modeling and  the effectiveness of the proposed control scheme
are validated on the benchmark quadrotor simulator  developed in \cite{2017Introduction}. Ttwo cases of experiments are conducted.  The main parameters of the QUAV are listed in Table~\ref{table_para}.

\begin{table}[htbp]  
	\caption{\label{tab:test}Main parameters of the QUAV}   
	\centering
	\begin{tabular}{llll}
		\toprule
		Param & Value & Units  & Definition \\
		\midrule
		\quad $m$ & 1.4 &  kg  & Mass \\
		\quad $g$ & 9.8  &   m/s$^2$ & Gravity \\
		\quad $J_{xx}$ & 0.0211 &   kg$\cdot$m$^2$ & Inertia moment\\
		\quad $J_{yy}$ & 0.0219 &   kg$\cdot$m$^2$ & Inertia moment\\
		\quad  $J_{zz}$ & 0.0366 & kg$\cdot$m$^2$ & Inertia moment\\
		\quad $d$ & 0.2250 & m & Fuselage radius\\
		\quad $C_t$ & $1.105e^{-5}$ & N/(rad/s)$^2$ &  Thrust coefficient\\
		\quad $C_m$ & $1.779e^{-7}$ & N$\cdot$m/(rad/s)$^2$ & Moment coefficient \\
		\bottomrule
	\end{tabular} 
	\label{table_para}
\end{table}

\begin{figure}[htbp]
	\begin{center}
		\includegraphics[scale=0.55]{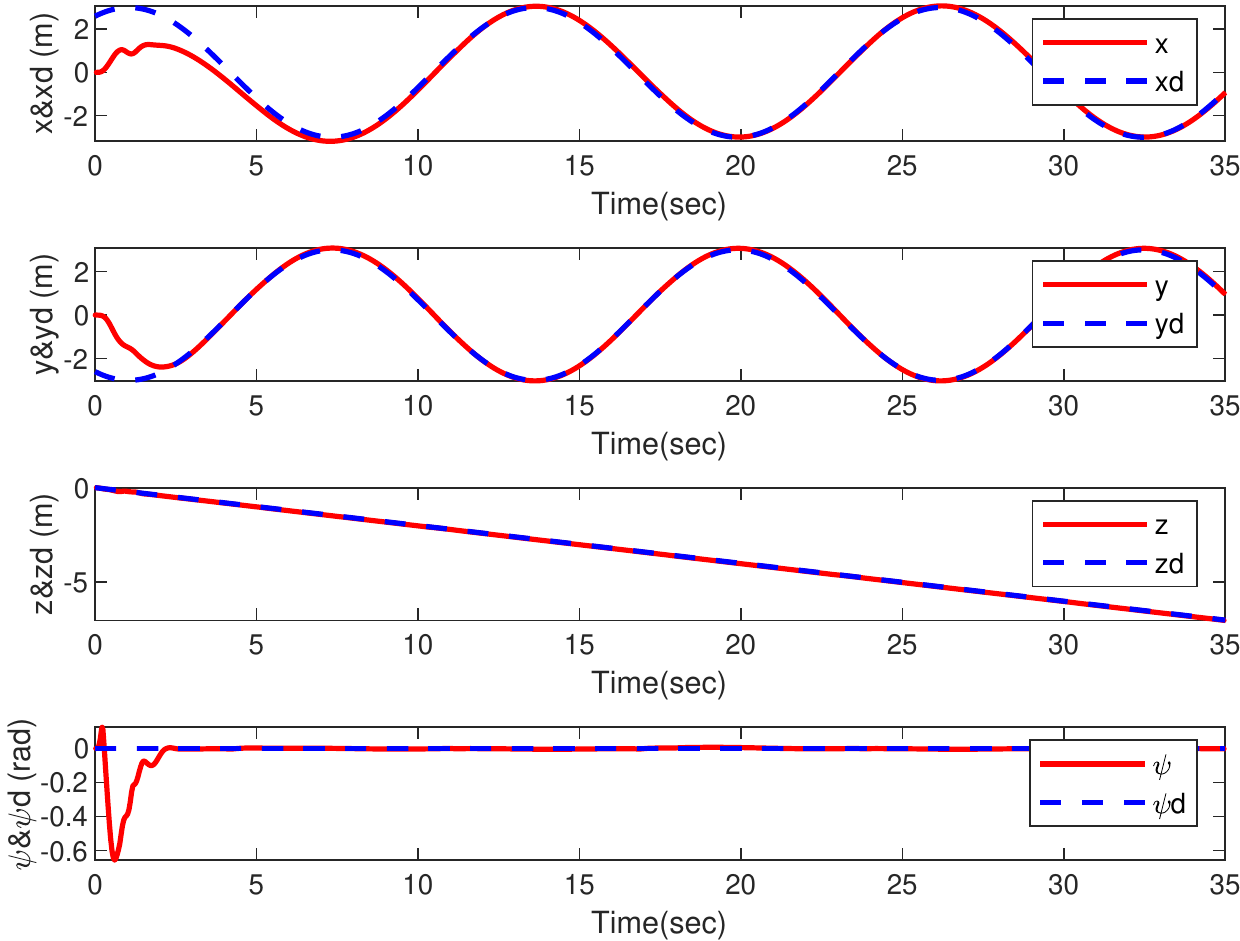}
		\par\end{center}
	\caption{Actual and desired trajectories $x,y,z$ and $\psi$ in Case 1 }
	\label{Case_1}
	\begin{center}
		\includegraphics[scale=0.45]{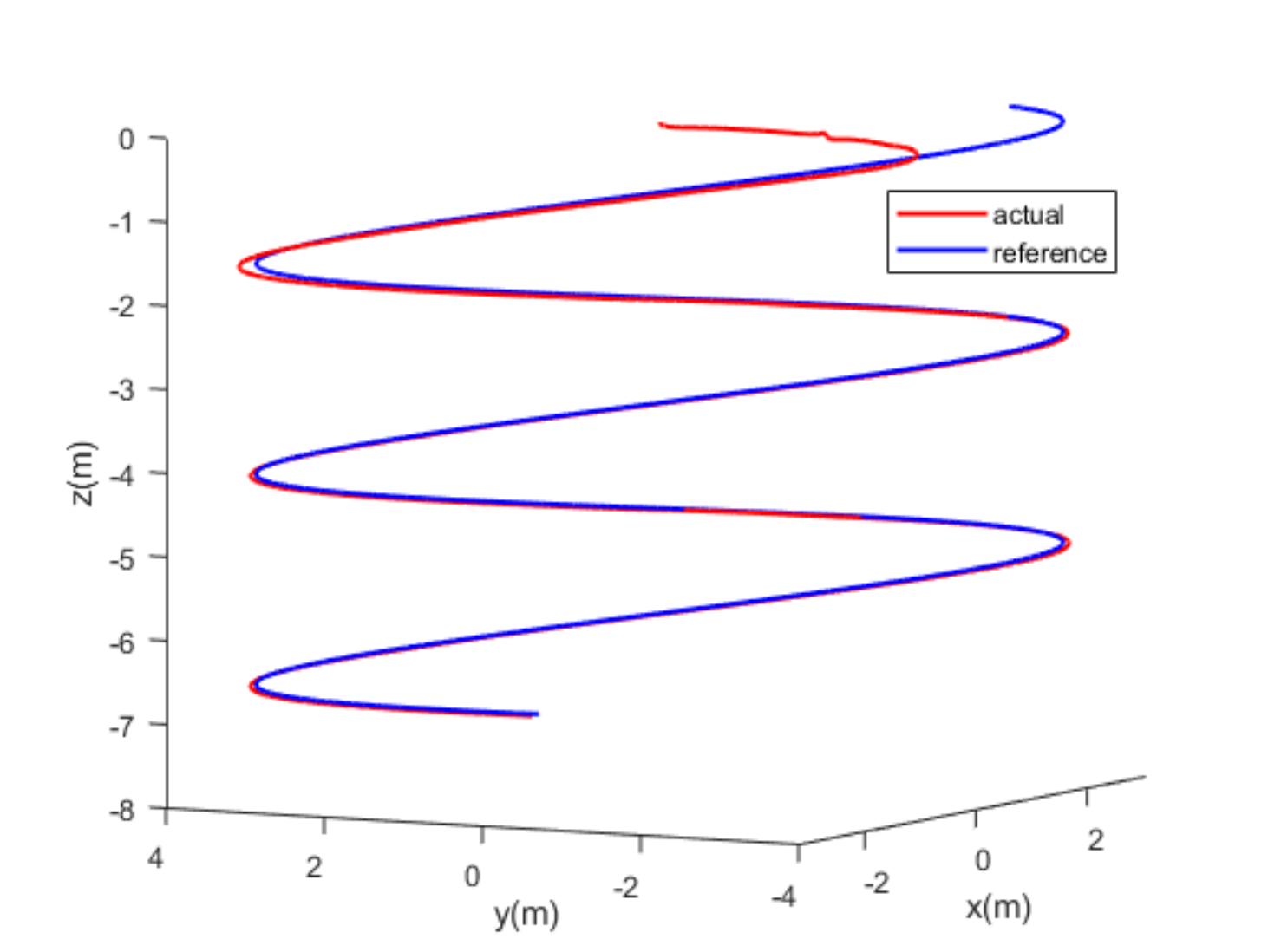}
		\par\end{center}
				\vspace{-4mm}
	\caption{Position tracking trajectories in 3-D space in Case 1 }
	\label{3D}
\end{figure}

 \text {Case \uppercase\expandafter{\romannumeral1}}: The reference trajectories of the position and yaw angle are chosen as $x_d(t)=3\sin(0.5t+\frac{\pi}{3})$, $y_d(t)=3\sin(0.5t-\frac{2\pi}{3})$,
 $ z_d(t)=-0.2t$, and $ \psi_d(t)=0$.
With the initial condition: $x(0)=0$, $y(0)=0$, $z(0)=0$, $\psi(0)=0.$ The backstepping gains are specified as
\begin{align*} 
	&c_1=11.52,c_2=28.00,c_3=16.63,c_4=6.54, \\
	&c_5=8.40,c_6=27.50,c_7=15.54,c_8=5.49.
\end{align*}
The position and yaw angle tracking trajectories are shown in Fig.~\ref{Case_1}, from which we can conclude  that the QUAV can track the references with satisfactory performance. Moreover, the actual and desired position trajectories in 3-D space are shown in Fig.~\ref{3D}, which also demonstrate the accuracy of the model and the effectiveness of associated controller.

\begin{figure}[htbp]
	\begin{center}
		\includegraphics[scale=0.55]{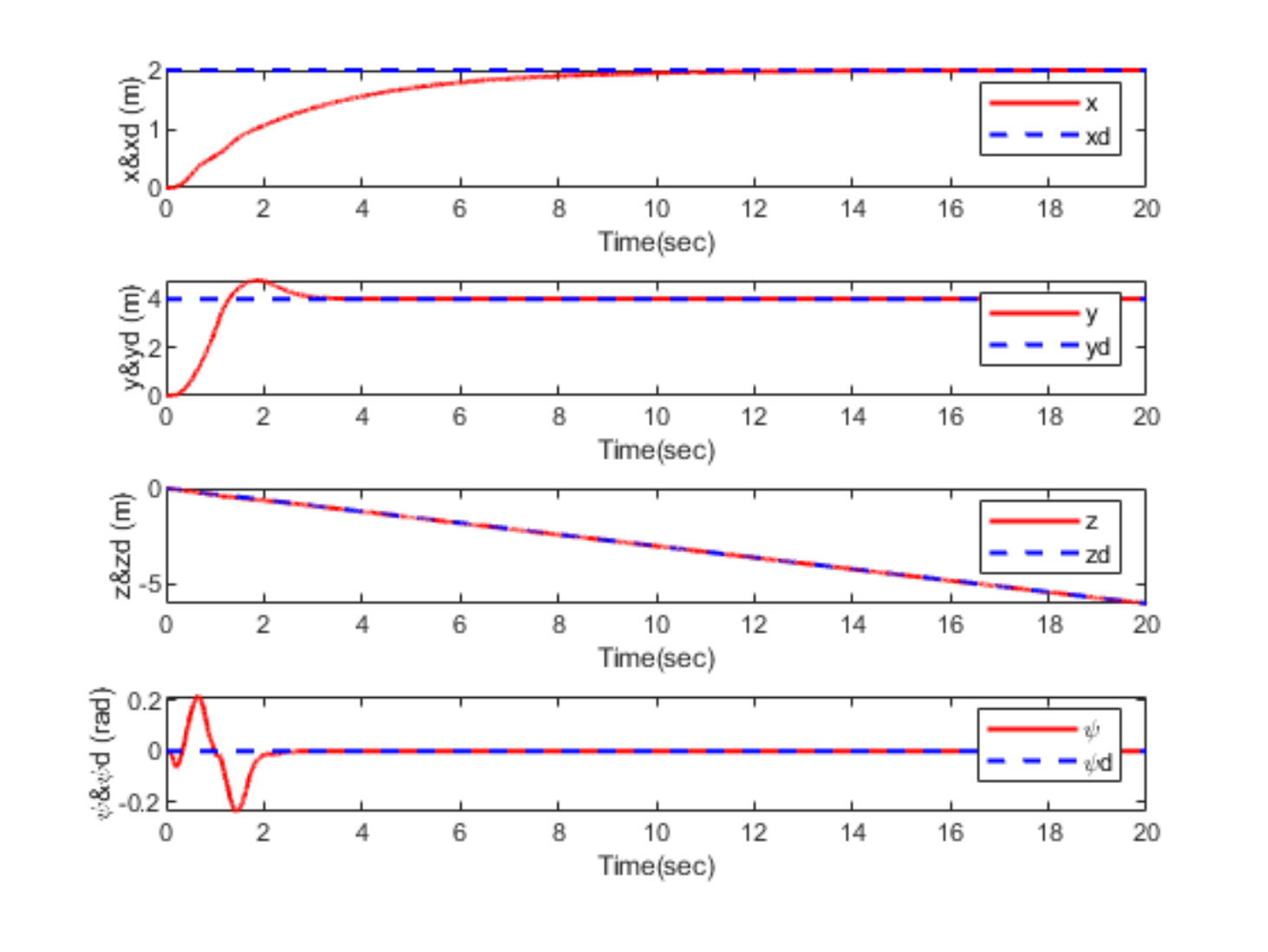}
		\end{center}
		\vspace{-8mm}
	\caption{Actual and desired trajectories $x,y,z$ and $\psi$ in Case 2 }
	\label{Case_2}
\end{figure}

\text {Case \uppercase\expandafter{\romannumeral2}}: The control objective in this case is to achieve  vertical take-off at a fixed horizontal position for the QUAV. The initial condition and the control gains are same as \text {Case \uppercase\expandafter{\romannumeral1}}. The reference trajectories are considered as
$x_d(t)=2$, $ y_d(t)=4 $, $z_d(t)=-0.3t$,  $\psi_d(t)=0.$
The experimental result  is presented in Fig.~\ref{Case_2}. It is observed that the tracking errors asymptotically converge to a small range within 4 seconds, and the final steady-state errors are approximately zero. 


\section{Conclusions}\label{six_section}
In this paper, we have proposed a novel modeling approach utilizing symbolic regression to construct the dynamic model of a QUAV. Based on the model constructed by symbolic regression, the PI and backstepping techniques were adopted to design a controller for time-varying position tracking. We have demonstrated the stability of the control system through  theoretical analysis. Subsequently, simulation studies have shown that the proposed control scheme can achieve superior tracking performance, and symbolic regression is especially effective. Future work can focus on considering external disturbance, e.g., \cite{chen2009remark}, and developing a robust control scheme based on symbolic regression.

\bibliographystyle{ieeetr}

 \fontsize{9}{9}
  \selectfont
\bibliography{bibfile}

\end{document}